\documentclass[a4paper]{jpconf}
\usepackage{graphicx}
\usepackage{color}
\usepackage{subfigure}
\usepackage{units}
\usepackage{booktabs}
\usepackage{amsmath}
\usepackage{amssymb}
\usepackage{feynmf}
\usepackage{textfit}
\usepackage{textcomp}
\usepackage[abs]{overpic}
\usepackage{ragged2e}
\usepackage{hyperref} 
\usepackage{tikz}
\usepackage{lineno}
\begin{document}


\title{Photon-hadron and photon-photon collisions in ALICE}

\author{Rainer Schicker, for the ALICE Collaboration}

\address{Physikalisches Institut, Im Neuenheimer Feld 226, 69120 Heidelberg}

\ead{schicker@physi.uni-heidelberg.de}

\begin{abstract}

A review is given on photon-hadron and photon-photon collisions in the ALICE 
experiment. The physics motivation for studying such reactions is outlined, 
and the results obtained in proton-lead and lead-lead collisions in Run 1 
of the LHC are discussed. The improvement in detector rapidity coverage due to 
a newly added detector system is presented. The ALICE perspectives for 
data taking in LHC Run II are summarised.     
  
\end{abstract}

\section{Introduction}

The ALICE experiment consists of a central barrel and of a forward muon 
spectrometer\cite{ALICE1}. Additional detectors are installed outside of the 
central barrel for trigger and event classification purposes. The excellent 
particle identification capabilities of the central barrel, in conjunction with
the low magnetic field, allow ALICE to investigate a variety of exclusive 
reaction channels as occuring in photon-hadron and photon-photon 
collisions\cite{ALICE2}. Here, the 
exclusivity of the particles produced within the central barrel can be imposed
 by requiring no signal in the detectors outside of the central barrel.
The ALICE collaboration has taken data from 2010 to 2013 in proton-proton, 
proton-lead and lead-lead collisions during Run I at the LHC. In the shut-down 
period 2013-2015 between Run I and Run II, a new detector system
ADA/ADC was installed. This new detector covers additional two units of 
pseudorapidity on both sides of the central barrel, and hence contributes
to an improved exclusivity condition of the analysed events, and to
an improved recognition of beam-gas background events.

\section{Photon-photon collisions}

Heavy-ion beams are the source of strong electromagnetic 
fields. In the Equivalent Photon Approximation (EPA), these fields are 
represented by an ensemble of equivalent photons\cite{Fermi,Budnev}. 
The cross section of heavy-ion induced photon-photon processes can be 
expressed as  

\begin{eqnarray}
\sigma^{EPA}_{PbPb \rightarrow PbPb\,X} = \int\int dn_{1,\gamma} \; dn_{2,\gamma} \; 
\sigma_{\gamma\gamma \rightarrow x}(\omega_1\,\omega_2),  
\label{eq:ggcross}
\end{eqnarray}

with dn$_{1,\gamma}$,\:dn$_{2,\gamma}$ the photon \mbox{flux in leading log. approximation of 
beam 1 and 2, respectively:}

\begin{eqnarray}
dn_{\gamma}(\omega,b) = \frac{Z^{2}\alpha}{\pi^{2}} 
\frac{d\omega}{\omega} \frac{d^{2}b}{b^{2}}.
\label{eq:ggflux}
\end{eqnarray}

In Eq. \ref{eq:ggflux}, the photon flux is given differentially in the photon 
energy $\omega$ and the impact parameter $b$ of the two colliding heavy ions. The 
photon-photon luminosity in EPA formalism is expressed in leading log. 
approximation as 

\vspace{-0.4cm}
\begin{eqnarray}
\frac{dL}{dWdy}\!\!=\!\!\frac{2}{W}
\frac{dL}{d\omega_1d\omega_2}\!\!=\!\!\frac{2}{W} \frac{dn_{1}(\omega_1)}{d\omega_1} 
\frac{dn_{2}(\omega_2)}{d\omega_2}\!\!=\!\!
\frac{4Z^{4}\alpha^{2}}{\pi^{2}(\omega_1\omega_2)^{3/2}} 
{\text{log}} (\frac{\gamma}{R\omega_1}) {\text{log}}(\frac{\gamma}{R\omega_2}),
\label{eq:gglumi}
\end{eqnarray}

with the invariant mass W=2$\sqrt{\omega_1\omega_2}$ and the rapidity 
y=$\frac{1}{2}$log$\big(\frac{\omega_1}{\omega_2}\big)$ of the two-photon 
system. Here, $\omega_{1}$ and $\omega_{2}$ denote the energies of the two 
interacting photons.

\begin{figure}[h]
\begin{center}
\begin{overpic}[width=.35\textwidth]{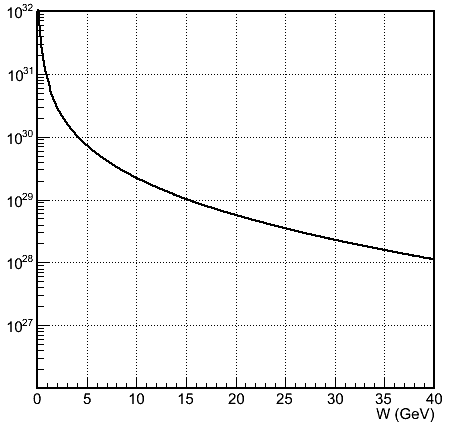}
\put(-2,28){\makebox(0,0){\rotatebox{90}
{\tiny L$_{\text{AA}}$dL/dWdy, cm$^{-2}$s$^{-1}$GeV$^{-1}$}}}
\put(12.,47.0){\small L$_{\text {PbPb}}$=4x10$^{26}$ cm$^{-2}$s$^{-1}$}
\end{overpic}
\end{center}
\vspace{-0.4cm}
\caption{Photon-photon luminosity for Pb-Pb at $\sqrt{s_{NN}}$ = 5.18 TeV.}
\label{fig1}
\end{figure}

\vspace{-0.2cm}
Shown in Figure \ref{fig1} is the photon luminosity 
$L_{AA} \frac{dL}{dWdy}$, calculated 
for a Pb-Pb luminosity of $L_{AA}$ = 4x10$^{26}$cm$^{-2}$s$^{-1}$
and a center-of-mass energy of $\sqrt{s_{NN}}$ = 5.18 TeV.

\section{Photon-hadron collisions}

The photons of the electromagnetic field of one nucleus can interact with 
a nucleus of the other beam, and can initiate a variety of photon-hadron 
reactions. Of particular interest here is the process of exclusive 
vector-meson photoproduction\cite{jcgn}.

\begin{figure}[h]
\begin{center}
\begin{minipage}[h]{0.34\textwidth}
\vspace{1.2cm}
\begin{overpic}[width=.9\textwidth]{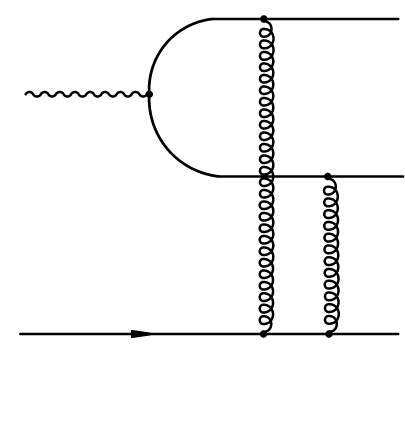}
\thicklines
\put(6.0,8.0){\small \it{p,Pb}}
\put(10.,42.){\small $\gamma^{*}$}
\put(26.,51.0){\small $q$}
\put(26.,32.){\small $\bar{q}$}
\end{overpic}
\end{minipage}
\hspace{.4cm}
\begin{minipage}[h]{0.51\textwidth}
\vspace{-0.4cm}
\begin{overpic}[width=.98\textwidth]{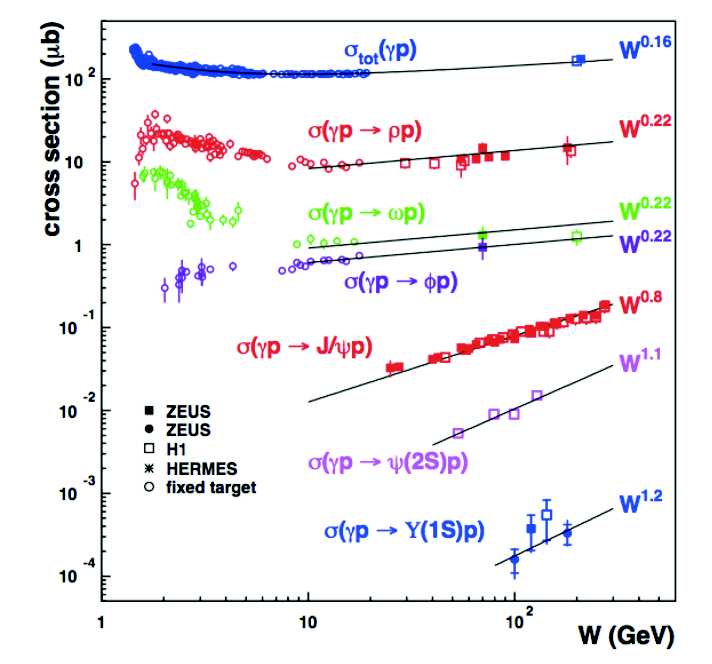}
\end{overpic}
\end{minipage}
\end{center}
\vspace{-.6cm}
\caption{Vector-meson photoproduction diagram on left, energy dependence 
of cross section on the right (Figure taken from \cite{vecmescross}).}
\label{fig2}
\end{figure}

The Feynman diagram for photoproduction of a vector-meson is shown in 
Figure \ref{fig2} on the left. The incoming photon fluctuates into a
$q\bar{q}$-excitation and gets scattered on-shell by the interaction. This 
interaction is due to colour-singlet exchange of gluons, hence is of 
diffractive nature. The systematic study of this process for vector-mesons 
of different masses, for low and high values of t (t\:=\:four momentum 
transfer) and for low and high Q$^{2}$ (Q$^{2}$\:=\:photon virtuality),
allows to study the nature of diffractive processes at a soft and hard scale,
and in the transition region in between. 
In the analysis of the $J/\psi$ cross section from HERA data, it was found 
that hard pomeron exchanges need to be included to get agreement with the data.
This hard pomeron contribution also significantly improves the Regge fits of 
the $\rho$-production cross section at high t-values \cite{DL_hardP}. 
Figure \ref{fig2} on the right shows the photoproduction cross section 
for vector-mesons of different masses as function of the photon-proton 
center-of-mass energy W. Clearly visible is the weak energy dependence
for the light mass of the $\rho$, and the  much steeper energy dependence
for the heavy masses of the  $J/\psi$ and $\Upsilon$.

The mass of the $J/\psi$ introduces a relatively hard scale, and hence
allows a perturbative QCD description. The amplitude can be factorized in
lowest order into the amplitude of the $\gamma \rightarrow c\bar{c}$
transition, the colour-singlet gluonic interaction of 
the $c\bar{c}$-system with the proton, and the evolution of the 
$c\bar{c}$-system into the $J/\psi$, resulting in a cross section of

\vspace{-0.5cm}
\begin{eqnarray}
\frac{d\sigma}{dt} (\gamma^{*}p \rightarrow J\!/\!\psi\; p) \big|_{t=0} = 
\frac{\Gamma_{ee}M_{J\!/\!\psi}^{3}\pi^{3}}{48 \alpha} 
\bigg[\frac{\alpha_{s}(\bar{Q}^{2})}{\bar{Q}^{4}} xg(x,\bar{Q}^{2})\bigg]^{2} 
\bigg( 1 + \frac{Q^{2}}{M_{J\!/\!\psi}^{2}} \bigg).
\label{eq:jpsicross}
\end{eqnarray}

\vspace{-0.2cm}
The cross section for $J/\psi$-photoproduction is shown 
in Eq. \ref{eq:jpsicross}. This cross section depends on the 
integrated gluon distribution $xg(x,Q^{2})$ squared, and hence 
allows the study of this gluon distribution at low x-values \cite{Martin}.
The analysis of this process on a Pb-nucleus gives access to the nuclear
gluon distribution, and is interesting for investigating nuclear medium 
efffects.  

\section{The ALICE experiment}

In the ALICE central barrel, the information from the Inner Tracking 
System (ITS), the Time Projection Chamber (TPC) and the Time-of-Flight 
system (TOF) allows momentum reconstruction and particle identification 
in the pseudorapidity range $-0.9 < \eta < 0.9$. A muon spectrometer 
covers the pseudorapidity range $-4. < \eta < -2.5$ \cite{ALICE1}. 

\vspace{-.2cm}
\begin{figure}[h]
\begin{center}
\begin{overpic}[width=.80\textwidth]{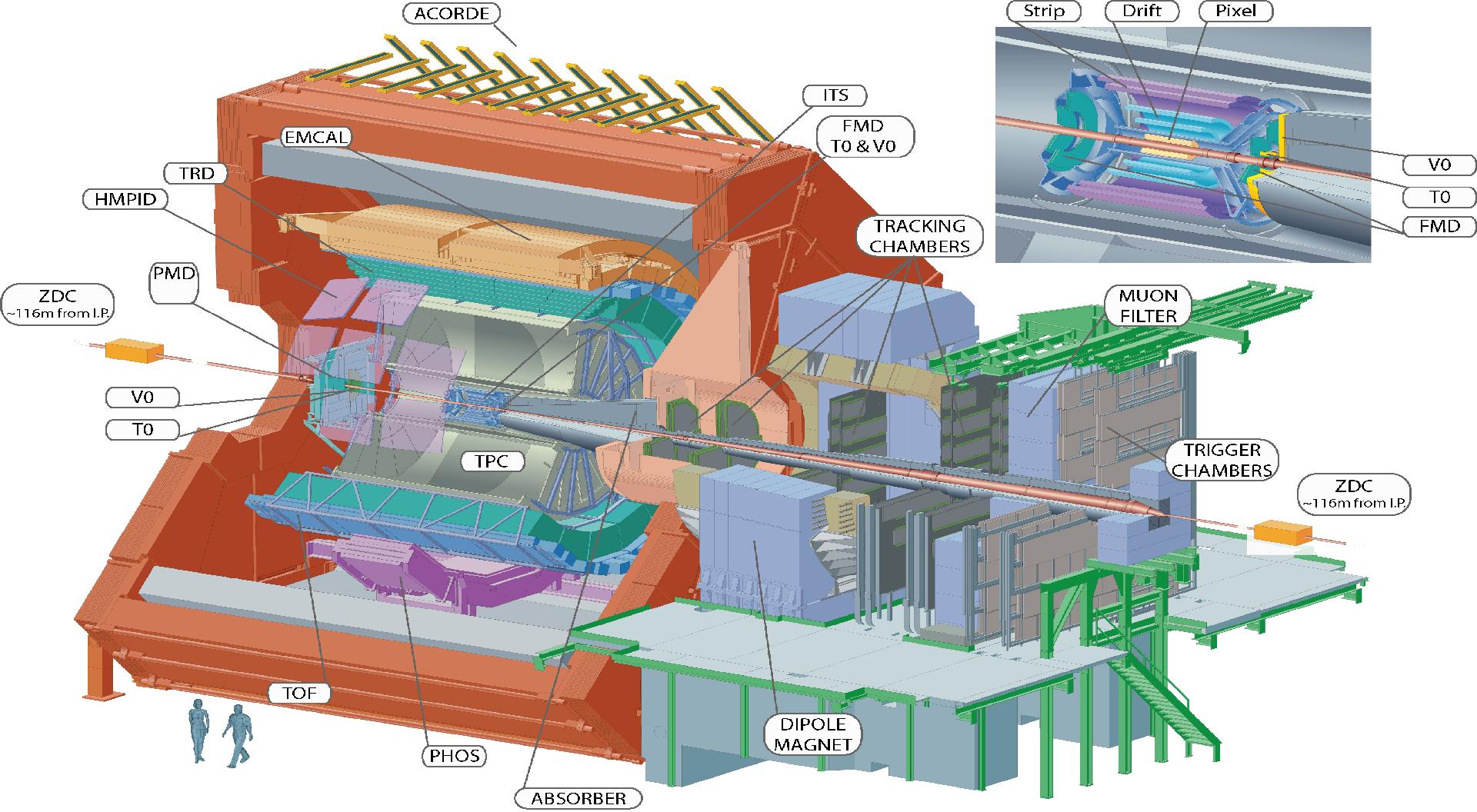}
\end{overpic}
\end{center}
\vspace{-.3cm}
\caption{The ALICE experiment.}
\label{fig3}
\end{figure}

\vspace{-.0cm}
The different detector systems of the ALICE experiment are displayed in 
Figure \ref{fig3}. The very forward neutral energy flow can be measured by 
Zero Degree Calorimeters (ZDC) located on both sides of the central barrel
at a distance of 116 m. Additional detectors are used for trigger and event 
classification purposes. First, the scintillator arrays VZERO-A and 
\mbox{VZERO-C} cover the range $2.8 < \eta < 5.1$ and $-3.7 < \eta < -1.7$, 
respectively. This detector is read out in 32 segments due to a four- and 
eightfold segmentation in pseudorapidity and azimuth, respectively.
Second, a Forward Multiplicity Detector (FMD) spans the 
range $1.7 < \eta < 5.1$ and $-3.4 < \eta < -1.7$, respectively.
Third, a new detector system ADA/ADC will be available for data
taking in Run II. This new system is installed on both
sides of the central barrel in the  pseudorapidity range 
$4.8 < \eta < 6.3$ and $-7. < \eta < -4.9$, respectively.

\section{$J/\psi$ production at forward rapidity in Pb-Pb collisions}

Exclusive photoproduction can be either coherent with the photon coupling
coherently to all nucleons, or incoherent with the photon coupling to
individual nucleons. Coherent production results in a low transverse 
momentum of $p_{T} \sim$ 60 MeV/c, with a 20-30 \% probability of 
breakup of the nucleus.
In incoherent production, the nucleus breaks up with a resulting
transverse momentum of the vector-meson $p_{T}$ of $\sim$ 500 MeV/c.

\begin{figure}[h]
\begin{center}
\begin{minipage}[h]{0.47\textwidth}
\begin{overpic}[width=.98\textwidth]{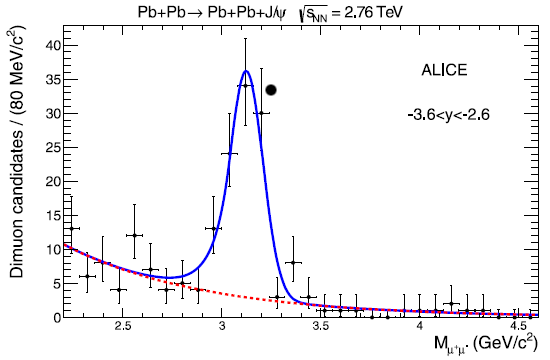}
\put(35.6,35.){\colorbox{white}{\makebox(1.0,1.0){}}}
\end{overpic}
\end{minipage}
\hspace{.8cm}
\begin{minipage}[h]{0.46\textwidth}
\begin{overpic}[width=.98\textwidth]{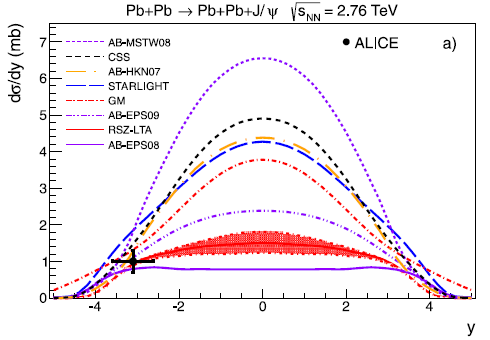}
\end{overpic}
\end{minipage}
\end{center}
\vspace{-0.4cm}
\caption{Invariant mass spectrum of two muons of opposite charge on the left, 
measured coherent cross section on the right with comparison to  
models (Figures taken from Ref. \cite{ALICE_jpsi_forw}).} 
\label{fig4}
\end{figure}

In Figure \ref{fig4}, the two track invariant mass spectrum of events 
measured in the ALICE muon spectrometer is shown \cite{ALICE_jpsi_forw}. 
These events, measured in the Pb-Pb run in the year 2011 with a dedicated 
trigger, contain exactly two muons of opposite charge, and correspond to 
an integrated luminosity of about 55$\mu b^{-1}$. The trigger used  here
consists of the requirements of a single muon trigger above
a $p_{T}$-threshold of 1 GeV/c, at least one hit in the VZERO-C detector
on the side of the muon spectrometer, and no hit in the VZERO-A
detector on the opposite side. 

The differential cross section for coherent $J/\psi$ production in the rapidity
\mbox{range $-3.6\!<\!y\!<\!-2.6$} derived from this data sample is 
$d\sigma/dy$ = 1.00 $\pm$ 0.18 (stat) $_{-0.26}^{+0.24}$(syst) mb. On the right 
hand side of Figure \ref{fig4}, this measured cross section is shown and 
compared to a number of model predictions. These models differ in the 
description of the photonuclear interaction. If no nuclear effects are taken 
into account, then all nucleons participate in the interaction, and the cross 
section is expected to scale with the number of nucleons squared, i.e. 
$\sigma \propto A^{2}$ (AB-MSTW08) \cite{AB}. A Glauber approach can be used to
calculate the number of nucleons which contribute to the interaction 
(STARLIGHT, GM, CSS) \cite{GM,CSS}. In this approach, the reduction in the 
calculated cross section depends on the $J/\psi$ cross section. In partonic 
models, the cross section is proportional to the square of the nuclear gluon 
distribution (AB-EPS08, AB-EPS09, AB-HKN07, RSZ-LTA) \cite{RSZ}.
The comparison of the measured cross section to the model calculations
shown on the right of Figure \ref{fig4} clearly shows that all these models 
give similar predictions at forward rapidity, and hence do not 
possess much discriminating power at this forward rapidity.

\section{$J/\psi$ production at midrapidity in Pb-Pb collisions}

During the Pb-Pb run in 2011, a dedicated trigger was running to select
events of exclusive production at midrapidity \cite{ALICE_jpsi_mid}. 
This trigger consisted of the requirements of at least two hits in the pixel 
detector of ITS, a number of fired pad-OR ($N^{on}$) in the TOF detector of 
$2 \leq N^{on} \leq 6$, and no hits in VZERO detector on either
side of the central barrel.    

\begin{figure}[h]
\begin{center}
\begin{minipage}[h]{0.38\textwidth}
\begin{overpic}[width=.98\textwidth]{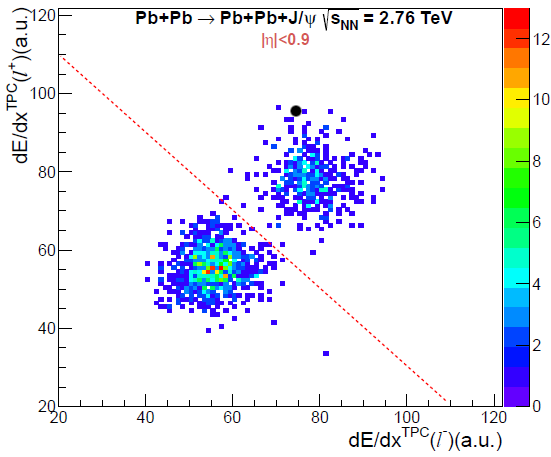}
\end{overpic}
\end{minipage}
\hspace{.2cm}
\begin{minipage}[h]{0.58\textwidth}
\begin{overpic}[width=.98\textwidth]{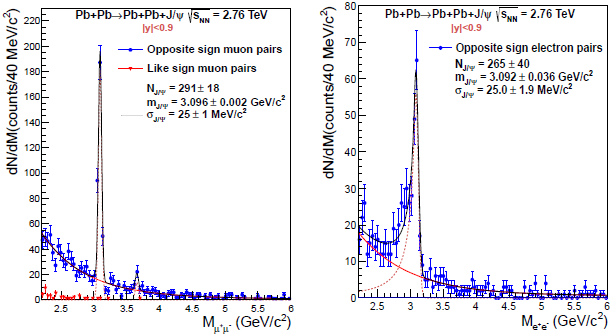}
\end{overpic}
\end{minipage}
\end{center}
\vspace{-0.3cm}
\caption{Electron and muon dE/dx signals in the TPC on the left, 
invariant masses for electron and muon pairs on the right for Pb-Pb
collisions at $\sqrt{s_{NN}}$=2.76 TeV
(Figures taken from Ref.\cite{ALICE_jpsi_mid}).} 
\label{fig5}
\end{figure}

In Figure \ref{fig5} on the left, the dE/dx signal in the ALICE TPC is 
plotted for lepton pairs which have an invariant mass in the range 
$2.8\leq M_{inv}\leq 3.2$ GeV/c$^{2}$. The signal of the negatively charged 
lepton of the pair is plotted on the horizontal axis, with the signal of 
the positively charged lepton on the vertical axis. The two signal areas 
of electron and muon pairs are clearly separated. On the right hand side of 
Figure \ref{fig5}, the invariant mass spectra are displayed for electron 
pairs with transverse momentum $p_{T} <$ 300 MeV/c, and for muon pairs with 
transverse momentum \mbox{$p_{T} <$ 200 MeV/c.} In the muon  channel, the 
peaks of the $J/\psi$ and $\psi$(2S) are clearly visible. 
The mass peak in the electron channel is asymmetric on the low mass side 
due to bremsstrahlung energy loss while passing through detector material. 

\begin{figure}[h]
\begin{center}
\begin{minipage}[h]{0.44\textwidth}
\begin{overpic}[width=.98\textwidth]{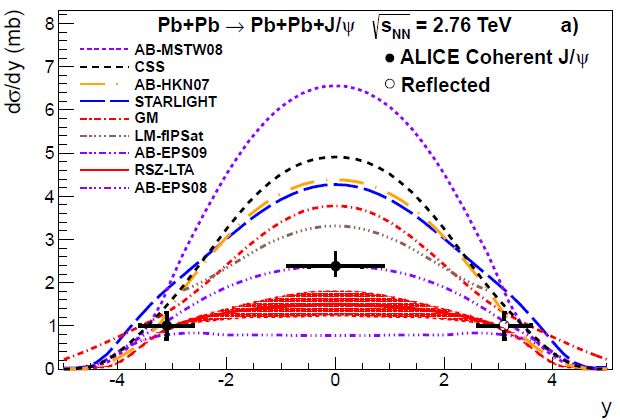}
\end{overpic}
\end{minipage}
\hspace{1.0cm}
\begin{minipage}[h]{0.42\textwidth}
\begin{overpic}[width=.98\textwidth]{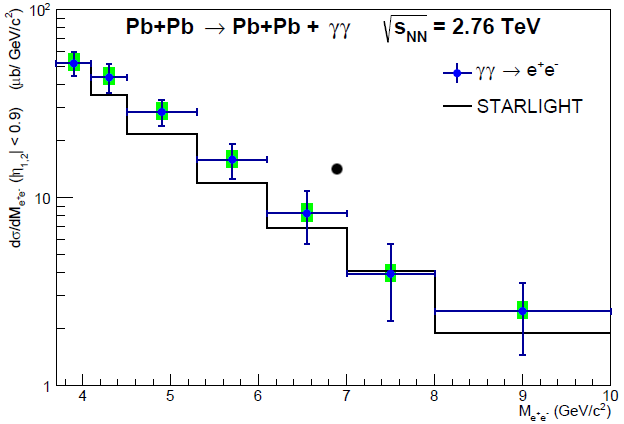}
\put(34.,26.){\colorbox{white}{\makebox(1.0,1.0){}}}
\put(43.,-.8){\colorbox{white}{\makebox(20.0,2.0)
{\tiny M$_{e^+e^-}$ (GeV/c$^2$)}}}
\put(-3.,5.){\colorbox{white}{\makebox(4.0,40.0){\rotatebox{90}
{\tiny d$\sigma\!/\!$dM$_{e^+e^-}$($|\eta_{1,2}|\!<\!0.9$)($\mu$\!b/GeV/c$^2$)}}}}
\end{overpic}
\end{minipage}
\end{center}
\vspace{-.3cm}
\caption{Cross section $J/\psi$ production at midrapidity on the left, 
cross section $\gamma\gamma \rightarrow$ e$^{+}$ e$^{-}$ on the right
(Figures taken from Ref.\cite{ALICE_jpsi_mid}).}
\label{fig6}
\end{figure}

The data sample taken with the trigger explained above corresponds to an 
integrated luminosity of 23 $\mu b^{-1}$.
From these data, the cross section value for coherent $J/\psi$ production at
midrapidity can be analyzed both in the electron and the muon decay channel,
yielding a combined value of $d\sigma_{J/\psi}^{coh}$/dy = 
2.38$_{-0.24}^{+0.34}$(stat + syst) mb as shown in Figure \ref{fig6} on the left. 
The combined value for the incoherent cross section is  
$d\sigma_{J/\psi}^{incoh}$/dy = 0.98$_{-0.17}^{+0.19}$(stat + syst) mb. 

From the spectrum shown in Figure \ref{fig5}, the two-photon cross section
$\gamma\gamma \rightarrow$ e$^{+}$e$^{-}$ can be extracted
by fitting the e$^{+}$e$^{-}$ continuum. This cross section is shown
on the right part of Figure \ref{fig6} for the mass range 
$3.7 \leq M_{inv} \leq 10$ GeV/c$^{2}$, and compared to STARLIGHT
predictions indicated by the black line \cite{STARLIGHT}.
The measured values are about 20\% higher than the STARLIGHT 
predictions, but are compatible within one standard deviation of the data.

\section{$\psi$(2S) production at midrapidity in Pb-Pb collisions}

The cross section for exclusive photoproduction of $\psi$(2S),
and comparison to the $J/\psi$ cross section, are of interest
since these two states are similar in mass but possess a different
radial wavefunction. It is expected that the radial node in the 
$\psi$(2S) wavefunction results in a reduced cross section.
The same data sample as described above was hence used to extract 
the coherent cross section for $\psi$(2S) production \cite{ALICE_jpsi2s_mid}.
The decay channels $\psi$(2S)  $\rightarrow l^{+} l^{-}$ can be extracted 
from the invariant lepton pair masses as shown in Figure \ref{fig5} above. 
In addition, the $\psi$(2S) can be identified by its decay 
$\psi$(2S) $\rightarrow J/\psi\: \pi^{+}\pi^{-}$ with subsequent identification 
of the $J/\psi$ in the lepton decay channels as explained above.

\begin{figure}[h]
\begin{center}
\begin{minipage}[h]{0.36\textwidth}
\begin{overpic}[width=.98\textwidth]{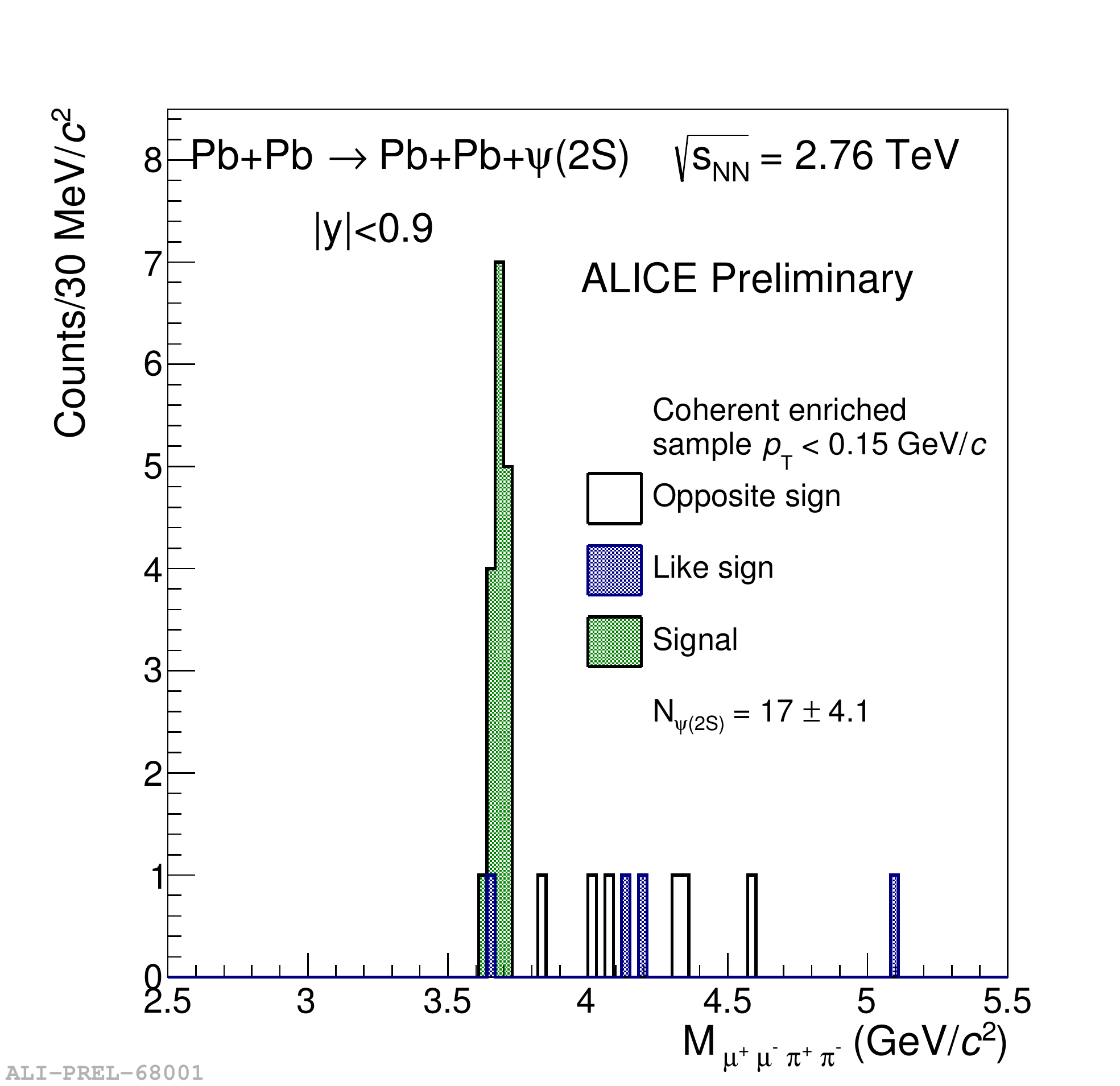}
\end{overpic}
\end{minipage}
\hspace{.8cm}
\begin{minipage}[h]{0.54\textwidth}
\begin{overpic}[width=.98\textwidth]{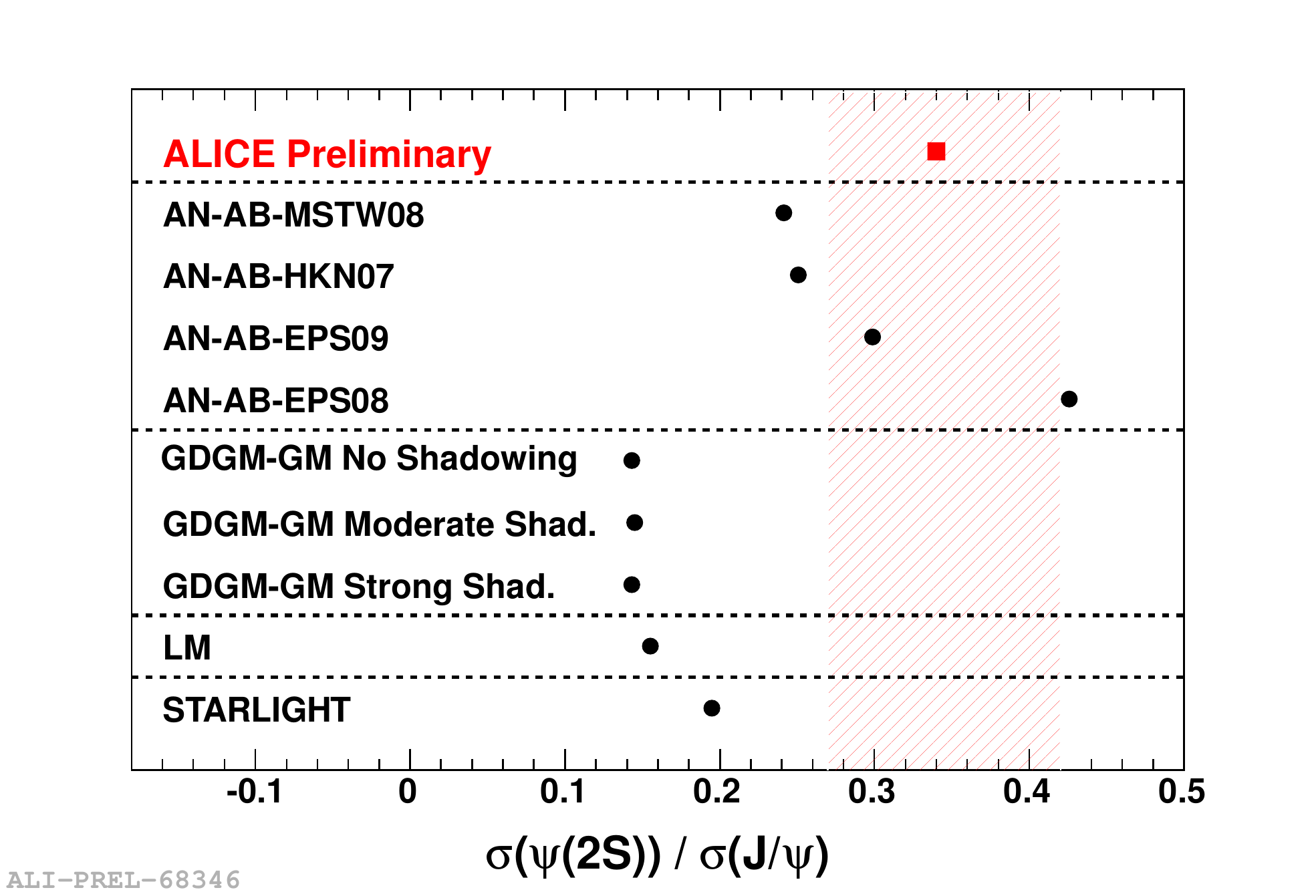}
\end{overpic}
\end{minipage}
\end{center}
\vspace{0.cm}
\caption{Invariant $\mu^{+}\mu^{-}\pi^{+}\pi^{-}$ mass spectrum on the left, 
comparison to models on the right
(Figures taken from Ref.\cite{ALICE_jpsi2s_mid}).}
\label{fig7}
\end{figure}

In Figure \ref{fig7} on the left, the invariant mass spectrum of 
$\mu^{+}\mu^{-}\pi^{+}\pi^{-}$ events is shown. Here, the invariant
mass of the muon pair has to meet the requirement of originating
from a $J/\psi$ decay.
A similar spectrum can be extracted from $e^{+}e^{-}\pi^{+}\pi^{-}$ events,
with the same condition on the electron pair mass of originating
from $J/\psi$ decay. From these events, and combined with the direct 
lepton decay channel,  a coherent $\psi$(2S) cross  section is derived 
for the rapidity interval $-0.9 <  y < 0.9$ of 
$d\sigma_{\psi(2S)}^{coh}/dy$ = 0.83$\pm$0.19 (stat+syst)mb.
 
The ratio of the cross section of $J/\psi$ to $\psi$(2S)
is of particular interest since many of the systematic uncertainties
cancel out. The ratio of these two cross sections is derived to be 
($d\sigma_{\psi(2S)}^{coh}/dy$)/($d\sigma_{J/\psi}^{coh}/dy$) = 
0.34$_{-0.07}^{+0.08}$ (stat+syst). This cross section ratio is shown in 
Figure \ref{fig7} on the right, and compared to  model predictions. 

The model by Adeluyi and Nguyen (AD) makes different 
predictions based on the different nuclear gluon PDF as defined 
in EPS08, EPS09, HKN07 and MSTW08 \cite{AN1,AN2,AN3,AN4}.
The model of Gay-Ducati, Griep and Machado (GDGM) is based 
on colour dipole model \cite{GDGM}, as well as the predictions
by Lappi and Mantysaari \cite{LM}. STARLIGHT uses Vector Dominance
model and a parameterization of existing HERA data \cite{Klein}.
Most of these models miss this cross section ratio by a factor
of 2-2.5. It is surprising that the AN model seems to describe
the data best, even though it assumes an identical wavefunction 
for the $J/\psi$ and $\psi$(2S) states.

\section{$\rho$-production at midrapidity in Pb-Pb collisions}

The study of $\rho$-photoproduction is interesting for a variety of reasons.
First, this reaction enables the study of a diffractive process at a soft
scale, a  region where perturbative QCD has little predictive power.
Data on $\rho$-photoproduction from the WA4 experiment at CERN can be well 
explained by a soft diffractive mechanism up to t-values of a few GeV$^{2}$, 
i.e. by soft pomeron and f$_{2}$,a$_{2}$-exchange \cite{WA4,Nachtmann}. 
At large t-values $t \sim$ 100 GeV$^{2}$, it is however found that
hard pomeron exchange is needed to explain the measured data \cite{DL_hardP}. 

\begin{figure}[h]
\begin{center}
\begin{minipage}[h]{0.36\textwidth}
\begin{overpic}[width=.98\textwidth]{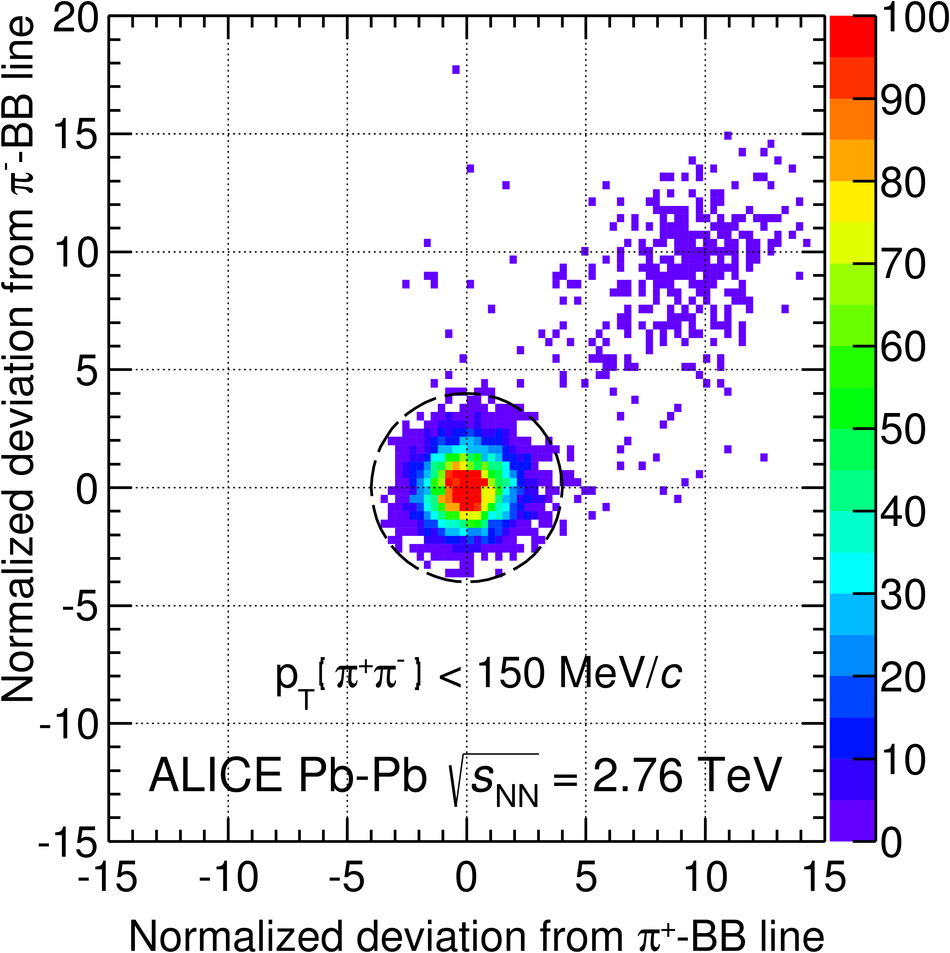}
\end{overpic}
\end{minipage}
\hspace{.8cm}
\begin{minipage}[h]{0.50\textwidth}
\begin{overpic}[width=.98\textwidth]{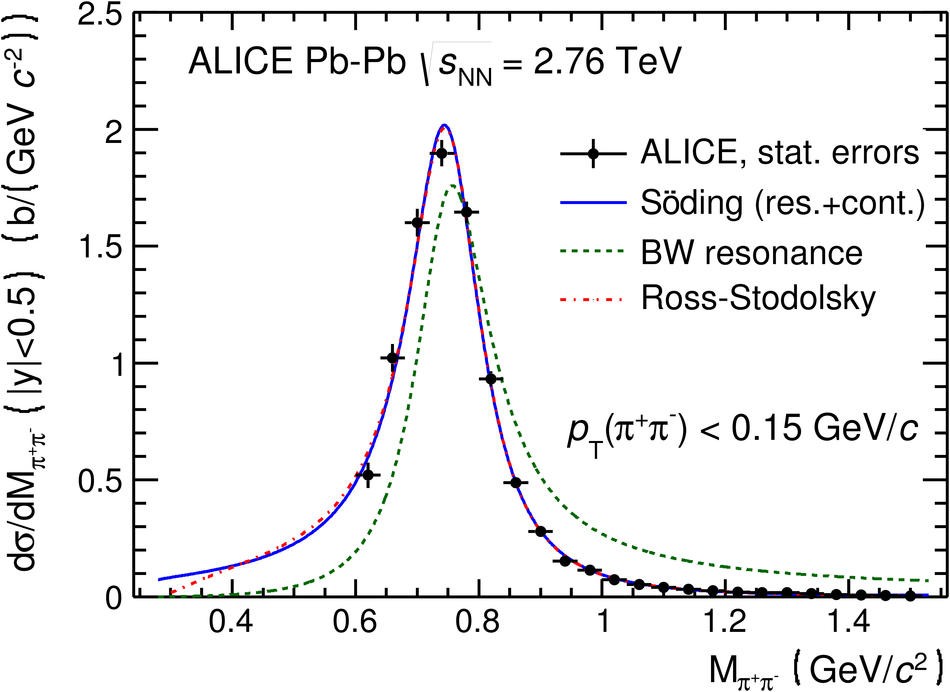}
\end{overpic}
\end{minipage}
\end{center}
\vspace{-.2cm}
\caption{Particle identification from TPC dE/dx on the left,
pion pair invariant mass spectra on the right
(Figures taken from Ref.\cite{ALICE_rho_mid}).}
\label{fig8}
\end{figure}

The data sample for analysing the $\rho$-photoproduction cross section 
was recorded by ALICE in the Pb-Pb run in 2010 at an energy of 
$\sqrt{s_{NN}}$ = 2.76 TeV \cite{ALICE_rho_mid}. With the low luminosity 
at the beginning of the run, a trigger requirement of at least two TOF hits 
was defined. With increasing luminosity, this requirement was refined to 
include additionally at least two hits from the pixel layer of the ITS, and 
no activity in the VZERO counters.  

In Figure \ref{fig8} on the left, the particle identification is shown 
derived from the dE/dx information of the TPC. 
On the horizontal axis, the deviation of the measured dE/dx value from the
expected dE/dx is shown for the positive pion, with the 
corresponding quantity for the negative pion on the vertical axis.
This deviation is normalized to units of standard deviation.
Clearly identified at the origin are $\pi^{+}\pi^{-}$ pairs, with
e$^{+}$e$^{-}$ pairs from photon-photon interactions located towards
the upper right corner. 

Figure \ref{fig8} on the right shows the invariant mass distribution of
$\pi^{+}\pi^{-}$ pairs. Here, a transverse momentum threshold of 
$p_{T} \leq$ 0.15 GeV/c is applied for selecting coherent $\rho$-production.  
A Breit-Wigner, a S\"{o}ding and a Ross-Stodolsky parameterization were used 
to fit the measured $\rho$-shape. A single Breit-Wigner parameterization 
shown in Figure \ref{fig8} by the green dashed line clearly cannot be used to 
describe the data. The S\"{o}ding parameterization, taken as the sum 
of a  Breit-Wigner and a continuum contribution, describes the measured data 
points well. The Ross-Stodolsky function, defined by a Breit-Wigner function 
multiplied by a mass dependent factor, describes the data  also well as 
indicated in Figure \ref{fig8} by the blue and red lines.

The analysis of the data taken with the two different triggers discussed 
above result in cross section values of 
11.8$\pm$1.6(stat)$_{-1.4}^{+1.1}$(syst)mb for the TOF only trigger, and 
9.4$\pm$0.7(stat)$_{-1.1}^{+0.9}$(syst)mb for the TOF-SPD-VZERO trigger. 
 
In the photoproduction process discussed here, additional photons
can be exchanged leading to electromagnetic excitation of the 
Pb-nucleus. This excitation may result in the forward emission of one 
or several neutrons measured by the ALICE ZDCs.
The STARLIGHT and RSZ model show good agreement with the measured 
neutron multiplicity distribution \cite{BKN,RSZ}.  

\section{$J/\psi$ production at forward rapidity in p-Pb collisions}

The ALICE collaboration has analysed $J/\psi$ photoproduction in the asymmetric
p-Pb system at a nucleon-pair centre-of-mass energy of $\sqrt{s_{NN}}$ = 
5.02 TeV at forward and backward \mbox{rapidities \cite{ALICE_jpsi_pPb}.} 
This process can occur on the proton $\gamma +$ p $\rightarrow J/\psi$ + p, 
or on the Pb-nucleus $\gamma +$  Pb $\rightarrow J/\psi +$Pb. The much 
larger photon flux of the Pb-nucleus, however, makes the production on 
the proton more likely. Such measurements are of high interest since 
the cross section is sensitive to the gluon PDF at low values of 
Bjorken-x as explained above. The search for signs of gluon saturation
at low-x is one of the open issues of studying QCD.
Such signs could, for example, be revealed by studying the cross
section of vector-meson photoproduction at the highest collider
energies now available, that is at the LHC.

\vspace{-0.2cm}
\begin{figure}[h]
\begin{center}
\begin{minipage}[h]{0.32\textwidth}
\begin{overpic}[width=.98\textwidth]{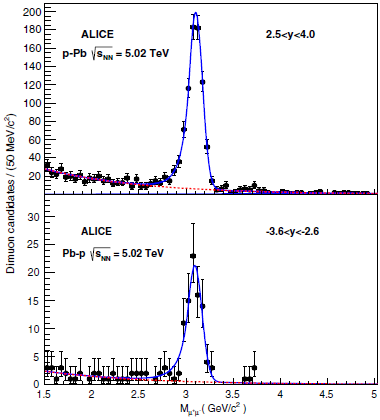}
\put(20.,-.8){\colorbox{white}{\makebox(20.0,2.0)
{\tiny M$_{\mu^+\mu^-}$ (GeV/c$^2$)}}}
\put(-3.,10.){\colorbox{white}{\makebox(4.0,40.0){\rotatebox{90}
{\tiny Dimuon candidates/(50 MeV/c$^2$) }}}}
\end{overpic}
\end{minipage}
\hspace{.8cm}
\begin{minipage}[h]{0.60\textwidth}
\begin{overpic}[width=.98\textwidth]{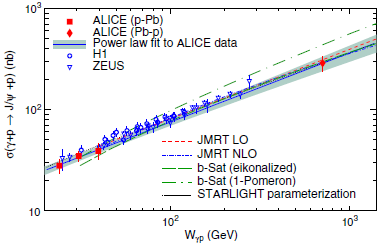}
\end{overpic}
\end{minipage}
\end{center}
\vspace{-.6cm}
\caption{Muon pair invariant mass on the left, $J/\psi$ photoproduction 
cross section on the right
(Figures taken from Ref.\cite{ALICE_jpsi_pPb}).}
\label{fig9}
\end{figure}

\vspace{-.2cm}
In Figure \ref{fig9} on the left, the invariant mass spectrum of
muon pairs measured in the muon spectrometer is shown.
The direction of the proton and lead beam were reversed in the LHC
to be able to measure at forward and backward rapidities, as shown
in the upper and lower part of this figure.
The measurement of the $J/\psi$ at positive rapidities corresponds
to a range of W$_{\gamma p}$ photon-proton center-of-mass energies
$21 < W_{\gamma p} < 45$ GeV, whereas negative rapidities
correspond to a range of $577 < W_{\gamma p} < 952$ GeV.
The muon pair invariant mass spectra at both forward and backward rapidity
clearly show a peak due to the $J/\psi$. The shape of the 
peak is well described by a Crystal Ball parameterization.
The dimuon continuum shown in Figure \ref{fig9} is fitted with 
an exponential form as expected to arise from two-photon 
continuum production. 

On the right side of Figure \ref{fig9}, the exclusive $J/\psi$ 
photoproduction cross section derived from the ALICE data is shown,
together with earlier cross section measurements and model
predictions. A power law fit to the data measured at HERA 
yields $\sigma \sim W_{\gamma p}^{\delta}$, with $\delta$ = 0.69 
and $\delta$ = 0.67  for the ZEUS and H1 data, respectively \cite{ZEUS,H1}.
An analogous fit to the ALICE cross section values yields
$\delta$ = 0.68$\pm$0.06 (stat+syst).
No significant deviation of the power law behaviour of the cross 
section is hence seen up to energies $W_{\gamma p} \sim$ 700 GeV. 
   
The ALICE data shown in Figure \ref{fig9} are compared to model predictions 
as follows: The  first model, labeled as JMRT LO, is based on a power law 
description of the process. The second, labeled JMRT NLO, includes next to 
leading order corrections\cite{JMRT}. Next, the b-Sat predictions are from an 
eikonalized model using the color glass condensate approach to include 
saturation\cite{bSat}. The STARLIGHT parameterization uses a power law fit 
based on fixed-target and HERA data, yielding $\delta$ = 0.65 $\pm$ 0.02. 
All these predictions agree to the measured data within one sigma of the 
quoted experimental uncertainty.
The b-Sat 1-Pomeron calculation from Ref. \cite{bSat1} is in agreement       
with the low energy ALICE data points, but is at odds at the high 
energy data points by about 4 sigmas.

\section{New detectors for RUN II}

In the shutdown period LS1 between Run I and Run II of the LHC, 
new detector systems ADA and ADC were installed on A and C-side
of ALICE, respectively. These detectors systems consist of double
layers of plastic scintillator paddles with light guides coupled to 
photomultipliers.

\begin{figure}[h]
\begin{center}
\begin{minipage}[h]{0.46\textwidth}
\begin{overpic}[width=.99\textwidth]{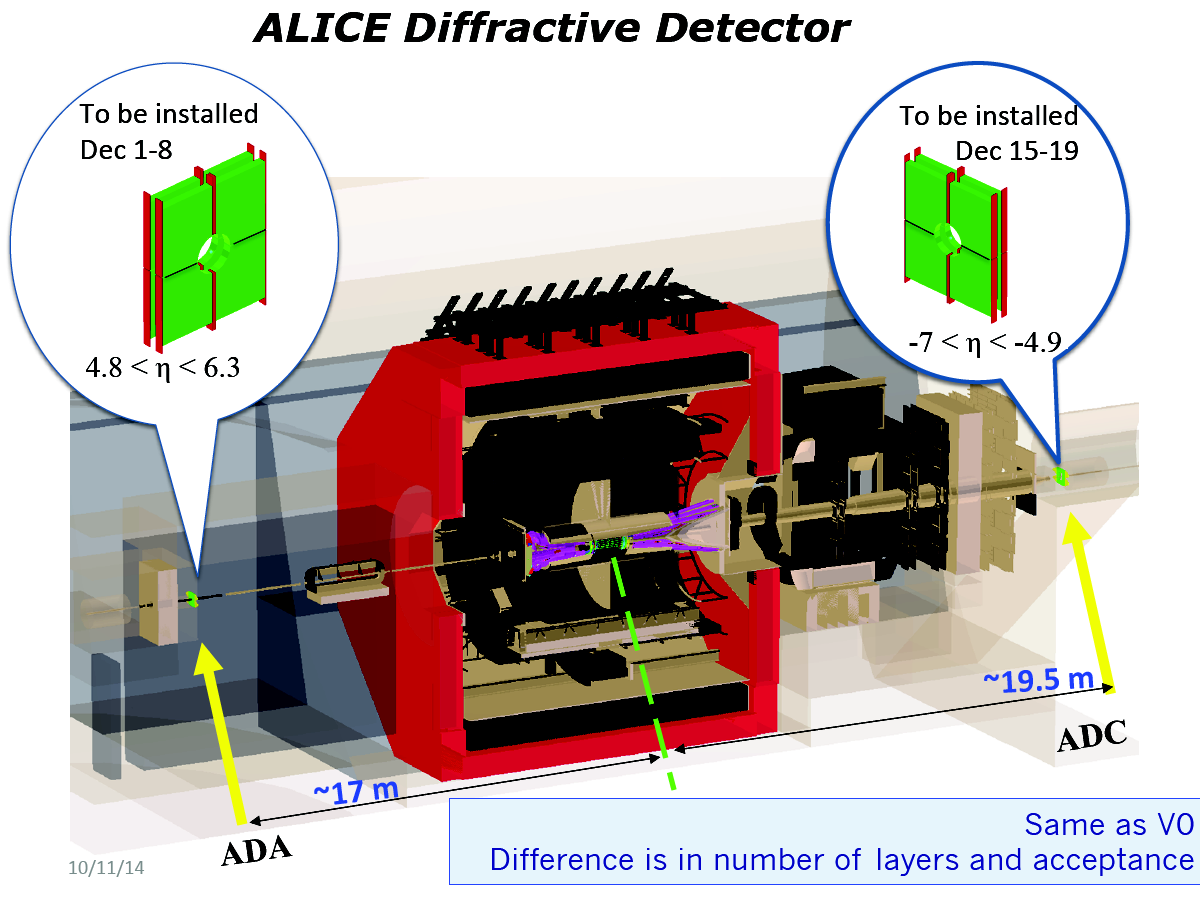}
\end{overpic}
\end{minipage}
\hspace{.4cm}
\begin{minipage}[h]{0.50\textwidth}
\begin{overpic}[width=.99\textwidth]{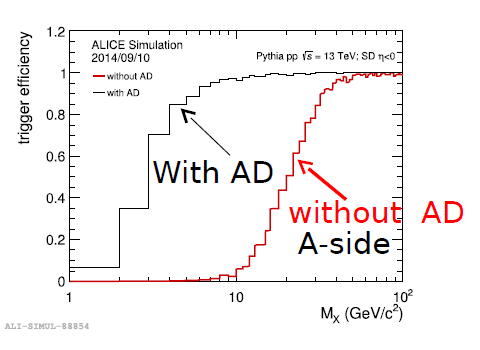}
\put(68.,20.){\colorbox{white}{\makebox(7.0,6.0){ }}}
\put(48.,15.){\colorbox{white}{\makebox(14.0,4.0){ }}}
\put(57.,17.5){\colorbox{white}{\makebox(4.0,2.0){\color{red}\bf AD}}}
\put(48.,14.){\colorbox{white}{\makebox(4.0,2.0){\bf A-side}}}
\end{overpic}
\end{minipage}
\end{center}
\vspace{0.cm}
\caption{Location of new detectors ADA and ADC in ALICE on the left,
trigger efficiency on A-side with and without AD as function of 
diffractive proton mass M$_{x}$ on the right.}
\label{fig10}
\end{figure}

In Figure \ref{fig10} on the left, the positions of these new
detector systems are shown. The ADA detector, located on A-side of ALICE,
covers the pseudorapidity range $4.8 < \eta < 6.3$, with ADC 
being located on C-side in the range $ -7 < \eta < -4.9$.
These two systems hence cover approximately two additional units
of pseudorapidity on both sides in very forward direction. 
On the right side of Figure \ref{fig10}, the acceptance on A-side is 
shown for diffractively excited protons of mass M$_{x}$. The threshold 
for 50\% efficiency is lowered from a diffracticve  mass of about 
\mbox{M$_{x} \sim$ 20 GeV/c$^{2}$} to a value of about M$_{x} \sim$ 
3 GeV/c$^{2}$.  These new detectors will improve the condition of 
exclusivity in future measurements of exclusive particle production, 
will extend the acceptance to lower diffractive masses, and will 
help in  rejection of beam-gas background events. 

\section{Plans for Run II}

Run II of the LHC has officially started with the declaration of stable
beams on June 3. The present center-of-mass energy in proton-proton
mode is $\sqrt{s}$ = 13 TeV. It is expected that the LHC will deliver
an integrated luminosity of 75-100 fb$^{-1}$ to the ATLAS and CMS experiments.
In the heavy-ion mode of Pb-Pb running, the center-of-mass energy will be 
$\sqrt{s_{NN}}$ = 5.1 TeV. The asymmetric system p-Pb will be measured either
at the energy of $\sqrt{s}$ = 5.1 or $\sim$\,8 TeV, corresponding
to the same energy of the Pb-Pb system, or to the highest energy possible.  

\section{Summary and outlook}

A wealth of information from Run I is available from the ALICE collaboration
regarding photon-hadron and photon-photon collisions.
The coherent and incoherent $J/\psi$ photoproduction cross section in Pb-Pb 
collisions has been  analysed at midrapidity and at forward rapidity.
The coherent $\psi$(2S) cross section has been  examined at midrapidity,
and the cross section ratio of these two resonances has been  given.
The $\rho$-photoproduction cross section has been presented, and the shape 
of the $\rho$ invariant mass spectrum has been studied. In addition, 
the $J/\psi$  photoproduction cross section has been  analysed at forward 
rapidity, and the $J/\psi$ photoproduction cross section has been examined
at midrapidity in the asymmetric p-Pb system.
The e$^{+}$e$^{-}$ invariant mass distribution resulting from photon-photon
interactions in Pb-Pb collisions  has been  studied at midrapidity. 
New detectors systems ADA and ADC have been installed for improved
detector coverage at forward rapidities for data taking in Run II.

\section{Acknowledgements}

This work is supported by the German Federal Ministry of Education and 
Research under promotional reference 05P15VHCA1.


\section*{References}
\begin{thebibliography}{9}

\bibitem{ALICE1}The ALICE Collaboration, JINST 3 (2008) S08002.

\bibitem{ALICE2}The ALICE Collaboration, Int.J.Mod.Phys. A29 (2014) 1430044. arXiv:1402.4476.

\bibitem{Fermi}E. Fermi, Z.Phys. A29, (1924), 315, doi:10.1007/BF03184853.

\bibitem{Budnev}V.M. Budnev, I.F. Ginzburg, G.V. Meledin, V.G. Serbo, 
Phys.Rept. 15 (1975) 181.  

\bibitem{jcgn}J.G. Contreras and J.D. Tapia Takaki, Int.J.Mod.Phys. A30
(2015) 1542012.

\bibitem{vecmescross}G. Wolf, Rept.Prog.Phys. 73 (2010), 116202, arXiv:0907.1217. 

\bibitem{DL_hardP}A. Donnachie and P.V. Landshoff, Phys.Lett.B478 (2000), 146, hep-ph/9912312. 

\bibitem{Martin}A.D. Martin, C. Nockles, M. Ryskin, T. Teubner, 
Phys.Lett. B662 (2008), 252, arXiv:0709.4406.  

\bibitem{ALICE_jpsi_forw}The ALICE Collaboration, Phys.Lett. B718 (2013) 1273, arXiv:1209.3715.

\bibitem{AB}A. Adeluyi and C.A. Bertulani, Phys.Rev. C85 (2012) 044904, arXiv:1201.0146.

\bibitem{GM}V.P. Goncalves and M.V.T. Machado, Phys.Rev. C84 (2011) 011902, arXiv:1106.3036.

\bibitem{CSS}A. Cisek, W. Sch\"{a}fer and A. Szczurek, Phys.Rev. C86 (2012) 014905, arXiv:1204.5381.

\bibitem{RSZ}V. Rebyakova, M. Strikman and M. Zhalov, Phys.Lett. B710, (2012) 647, 
arXiv:1109.0737. 

\bibitem{ALICE_jpsi_mid}The ALICE Collaboration, Eur.Phys.J.C73 (2013) 11, 2617, arXiv:1305.1467.  

\bibitem{STARLIGHT}http://starlight.hepforge.org/.

\bibitem{ALICE_jpsi2s_mid}The ALICE Collaboration, acccepted for publication in Physics Letters B, arXiv:1508.05076.

\bibitem{AN1}K.J. Eskola, H. Paukkunen and C.A. Salgado, JHEP 0807 (2008) 102, arXiv:0802.0139.

\bibitem{AN2}K.J. Eskola, H. Paukkunen and C.A. Salgado, JHEP 0904 (2009) 065, arXiv:0902.4154.

\bibitem{AN3}M. Hirai, S.Kumano and T.-H. Nagai, Phys.Rev. C76 (2007) 065207, arXiv:0709.3038.

\bibitem{AN4}A.D. Martin, W.J. Stirling, R.S. Thorne and G.Watt, Eur.Phys.J.C63 (2009) 189, arXiv:0901.0002.  

\bibitem{GDGM}M.B. Gay Ducati, M.T. Griep and M.V.T. Machado, Phys.Rev. C88 (2013) 014910, arXiv:1305.2407. 

\bibitem{LM}T. Lappi and H. Mantysaari, Phys.Rev. C87 (2013) 3, 032201, arXiv:1301.4095.

\bibitem{Klein}S. Klein and J. Nystrand, Phys.Rev. C60 (1999) 014903, hep-ph/9902259.

\bibitem{WA4}D. Aston et al., Nuclear Physics B209 (1982) 56. 

\bibitem{Nachtmann}A. Donnachie, H.G. Dosch, P.V. Landshoff and O. Nachtmann, 
{\it Pomeron Physics and QCD (Cambridge University Press, 2002)}.

\bibitem{ALICE_rho_mid}The ALICE Collaboration, JHEP 1509 (2015) 095, arXiv:1503.09177.

\bibitem{BKN}A.J. Baltz, S.R. Klein and J. Nystrand, Phys.Rev.Lett. 89 (2002) 012301, nucl-th/0205031.  

\bibitem{ALICE_jpsi_pPb}The ALICE Coll., Phys.Rev.Lett. 113 (2014) 23, 232504, arXiv:1406.7819.

\bibitem{ZEUS}ZEUS Collaboration, Eur.Phys.J. C24 (2002) 345, hep-ex/0201043.

\bibitem{H1}H1 Collaboration, Eur.Phys.J. C73 (2013) 6, 2466, arXiv:1304.5162.

\bibitem{JMRT}S.P. Jones, A.D. Martin, M.G. Ryskin, T. Teubner, JHEP 1311 (2013) 085, arXiv:1307.7099.

\bibitem{bSat}H. Kowalski, L. Motyka and G. Watt, Phys.Rev. D74 (2006) 074016.

\bibitem{bSat1}J.L.Abelleira Fernandez et al., arXiv:1211.4831.

\end{thebibliography}
\end{document}